\documentstyle[12pt]{article}
\newcommand{\newsection}{    
\setcounter{equation}{0}
\section}
\newcommand{\tr}[1]{\,{\rm tr}\,#1}
\def\e{{\,\rm e}\,}

\def\eop{\vspace*{\fill}\pagebreak}
\def\be{\begin{equation}}
\def\ee{\end{equation}}
\def\bea{\begin{eqnarray}}
\def\eea{\end{eqnarray}}

\newcommand{\rf}[1]{(\ref{#1})}
\newcommand{\eq}[1]{Eq.~(\ref{#1})}

\def\L{\Lambda}
\def\l{\lambda}

\def\om{\omega}

\def\appendix#1{
  \addtocounter{section}{1}
  \setcounter{equation}{0}
  \renewcommand{\thesection}{\Alph{section}}
  \section*{Appendix \thesection\protect\indent #1}
  \addcontentsline{toc}{section}{Appendix \thesection\ \ \ #1}
  }
\newcommand{\ie}{{\it i.e.}\ }

\renewcommand{\d}{{{\partial}}}
\newcommand{\p}{{\prime}}
\newcommand{\ra}{\rightarrow}
\newcommand{\fr}[2]{{\textstyle {#1 \over #2}}}

\begin{document}

\begin{titlepage}
\begin{flushright}
ITEP-YM-2-93 \\ March, 1993
\end{flushright}
\vspace{1cm}

\begin{center}
{\LARGE An Exact Solution of Induced Large-$N$ \\ \vspace{0.6cm}
Lattice Gauge Theory at Strong Coupling} \end{center} \vspace{.5cm}
\begin{center}
{\large Yu.\ Makeenko}\footnote{E--mail: \ makeenko@nbivax.nbi.dk \ / \
makeenko@desyvax.bitnet \ / \ makeenko@vxitep.itep.msk.su \ }
\\ \mbox{} \\
{\it Institute of Theoretical and Experimental Physics} \\ {\it
B.Cheremuskinskaya 25, 117259 Moscow, RF}
\end{center}

\vspace{1cm}

\begin{abstract}
I show that the strong coupling solution of the Kazakov--Migdal model
with a general interaction potential $V(\Phi)$ in $D$ dimensions
coincides at large $N$ with that of the hermitean one-matrix model
with the potential $\tilde{V}(\Phi)$: $$
(2D-1)\tilde{V}^\prime = (D-1)V^\prime + D\sqrt{(V^\prime)^2+4(1-2D)\Phi^2},
$$ whose solution is known. The proof is given for an even potential
$V(\Phi)=V(-\Phi)$ by solving loop equations.
\end{abstract}

\vspace{2.5cm}
\noindent
Submitted to {\sl Modern Physics Letters A}

\eop
\end{titlepage}
\setcounter{page}{2}

\newsection{Introduction}

Solvable matrix models are usually associated with $D\leq1$ dimensional
theories. Kazakov and Migdal~\cite{KM92} have recently proposed
that the model (KMM) defined by the partition function
\be
Z_{KMM}=\int \prod_{x,\mu} dU_{\mu}(x) \prod_x
 d\Phi_x \e^{ \sum_x N_c \tr{}\left(-V(\Phi_x)+
\sum_{\mu=1}^D\Phi_x U_\mu(x)\Phi_{x+\mu}U_\mu^\dagger(x)\right)}\,,
\label{spartition}
\ee
where the scalar field $\Phi_x$ is in the
adjoint representation of the gauge group $SU(N_c)$ and the link variable
$U_\mu(x)$ belongs to the gauge group,
is solvable in the limit of large number of
colors, $N_c$, for $D>1$. Since stringy phenomena are associated with  the
strong coupling phase (see~\cite{Kaz92} for a review),
one is interested in the solution of KMM at strong coupling.

The original idea to solve KMM is based on the fact
that the scalar field can be diagonalized by a (local) gauge
transformation so that only ${\cal O}(N_c)$ degrees of freedom are left and
the saddle-point method is applicable as $N_c\ra\infty$. Migdal~\cite{Mig92a}
proposed to solve the saddle point equation by the
Riemann--Hilbert method and derived a master field equation to determine the
$N_c=\infty$ solution in the strong coupling phase. An explicit solution of
this equation for the quadratic potential was found by
Gross~\cite{Gro92}. A surprising property of the master field equation (not
yet completely understood) is that it admits~\cite{Mig92a,Mig92c}
self-consistent scaling solutions with non-trivial critical indices.

There are two other approaches to solving KMM at strong coupling.
The first one is based on loop equations which
were derived and solved for KMM with the quadratic potential by the
author~\cite{Mak92} recovering the solution of Ref.~\cite{Gro92}.
The second one,  proposed by Boulatov~\cite{Bou92b}, relies on the
relation of KMM to a matrix model on the Bethe lattice, whose ``naive''
 continuum limit is equivalent for $D>1$ to
a one-matrix model with the upside-down potential.

In the present paper I extend the approach based on loop equations
to the case of KMM with an
arbitrary potential. On the one hand, this provides a way to solve the model
at strong coupling which is an alternative to the Riemann--Hilbert method. On
the other hand, this method is along the line of modern studies of matrix
models of 2D quantum gravity by means of loop equations (for a review,
see Ref.~\cite{Mak91}).

The main results of this paper are as follows.
The loop equations are drastically simplified in the strong coupling phase at
$N_c=\infty$ due to the fact that
the averages of closed Wilson loops vanish except
for the loops with vanishing minimal area. The resulting equations are
satisfied for an even potential $V(\Phi)=V(-\Phi)$
by the ansatz which reduces KMM in $D$ dimensions to a hermitean
{\it one-matrix\/} model with the potential
 $\tilde{V}(\Phi)$:
\be
\tilde{V}^\prime(\Phi) = \frac{D-1}{2D-1}V^\prime(\Phi)
+ \frac{D}{2D-1}\sqrt{(V^\prime(\Phi))^2+4(1-2D)\Phi^2}
\label{newpotential}\,.
\ee
The solution of this model with an arbitrary potential is well-known.

In Section~2 I derive the exact loop equations for KMM with an arbitrary
potential and any $N_c$. In Section~3 I show how these equations are simplified
in the strong coupling phase at $N_c=\infty$. In Section~4 I define the ansatz
and obtain the exact solution of the $N_c=\infty$ loop equations in the case
of the even potential.
Section~5 is devoted to the explanation of the form of the solution
from the viewpoint of
the large mass expansion. In Section~6 I discuss some properties
of the solution and show how it can be compared to the one obtained
by the Riemann--Hilbert method.
Appendix~A contains the details of derivation of the loop equations.
In Appendix~B I analyze the large mass expansion of the one-link correlator
of the gauge fields at $N_c=\infty$.

\newsection{Loop equations for arbitrary potential \label{loops}}

The loop equations of KMM relate the closed adjoint Wilson
loops
\be
W_A(C)=\left\langle\frac{1}{N_c^2}
\left( \left| \tr{U(C)}\right|^2-1\right)\right\rangle
\label{adjloop}
\ee
to the open ones with the matter field attached at the ends:
\be
G_{\l}(C_{xy})= \left\langle
\frac {1}{N_c} \tr{}{\Big( \Phi_x U (C_{xy})
\frac{1}{\l- \Phi_y}
U^\dagger(C_{xy}) \Big)} \right\rangle\,.
\label{sG}
\ee

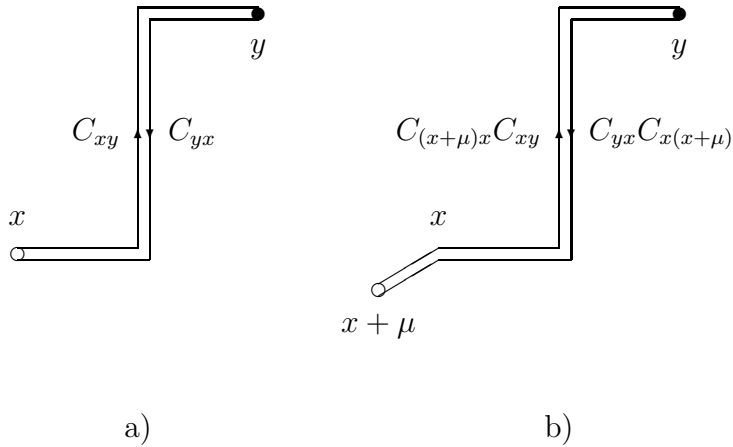
\begin{figure}[tbp]
\begin{picture}(120.00,190.00)(-34,20)
\unitlength=0.80mm
\linethickness{0.5pt}
\put(30.00,50.00){\line(3,0){20.00}}
\put(50.00,90.00){\line(3,0){20.00}}
\put(100.00,50.00){\line(3,0){20.00}}
\put(120.00,90.00){\line(3,0){20.00}}
\put(30.00,55.00){\makebox(0,0)[cc]{$x$}}
\put(70.00,83.00){\makebox(0,0)[cc]{$y$}}
\put(100.00,55.00){\makebox(0,0)[cc]{$x$}}
\put(140.00,83.00){\makebox(0,0)[cc]{$y$}}
\put(90.00,44.00){\line(5,3){10.00}}
\put(90.00,37.00){\makebox(0,0)[cc]{$x+\mu$}}
\put(50.00,20.00){\makebox(0,0)[cc]{a)}}
\put(120.00,20.00){\makebox(0,0)[cc]{b)}}
\put(50.00,50.00){\vector(0,1){20.00}}
\put(50.00,90.00){\line(0,-3){20.00}}
\put(120.00,50.00){\vector(0,1){20.00}}
\put(120.00,70.00){\line(0,3){20.00}}
\put(30.00,49.00){\circle{2.00}}
\put(70.00,89.00){\circle*{2.00}}
\put(52.00,88.00){\vector(0,-1){20.00}}
\put(52.00,48.00){\line(0,3){20.00}}
\put(30.00,48.00){\line(3,0){22.00}}
\put(52.00,88.00){\line(3,0){18.00}}
\put(90.00,43.00){\circle{2.00}}
\put(140.00,89.00){\circle*{2.00}}
\put(122.00,88.00){\line(3,0){18.00}}
\put(122.00,88.00){\vector(0,-1){20.00}}
\put(122.00,48.00){\line(0,3){20.00}}
\put(100.00,48.00){\line(3,0){22.00}}
\put(90.00,42.00){\line(5,3){10.00}}
\put(55.00,69.00){\makebox(0,0)[lc]{$C_{yx}$}}
\put(47.00,69.00){\makebox(0,0)[rc]{$C_{xy}$}}
\put(125.00,69.00){\makebox(0,0)[lc]{$C_{yx}C_{x(x+\mu)}$}}
\put(117.00,69.00){\makebox(0,0)[rc]{$C_{(x+\mu)x}C_{xy}$}}
\end{picture}
\caption[x]   {\hspace{0.2cm}\parbox[t]{13cm}
{\small
   The graphic representation for $G_\l(C_{xy})$ (a)
   and $G_\l(C_{(x+\mu)x}C_{xy})$ (b) entering \eq{sd}. The
   empty circles represent $\Phi_x$ or $\Phi_{x+\mu}$ while the filled
   ones represent $\frac{1}{\l-\Phi_y}$.
   The oriented solid lines represent the
   path-ordered products $U(C_{xy})$ and $U(C_{(x+\mu)x}C_{xy})$.  The color
   indices are contracted according to the arrows.}}
\label{fig.4}
\end{figure}

The loop equations result from the invariance of the measure in
\eq{spartition} under an arbitrary shift of $\Phi$ and read
\bea
 \left\langle \frac{1}{N_c} \tr{}\Big(V^\p(\Phi_x)
U(C_{xy}) \frac{1}{\l-\Phi_y}
U^\dagger(C_{xy}) \Big) \right\rangle
-\sum_{\mu=-D\atop \mu\neq0}^D G_\l(C_{(x+\mu)x}C_{xy})
  \nonumber \\* = \delta_{xy}
\left\langle \frac{1}{N_c} \tr{}{\Big( U(C_{xy})\frac{1}{\l-\Phi_y}\Big)}
\frac{1}{N_c} \tr{}{ \Big(\frac{1}{\l-\Phi_y}
U^\dagger(C_{xy})\Big)}\right\rangle
\label{sd}
\eea
where the path $C_{(x+\mu)x}C_{xy}$
on the l.h.s.\ is obtained by attaching the link $(x,\mu)$ to the path
$C_{xy}$ at the end point $x$ as is depicted in Fig.~\ref{fig.4}.
The details of derivation are presented in Appendix~A.
I have omitted here and below additional contact
terms which arise at finite $N_c$ due to the fact that $\Phi$ belongs to
the adjoint representation, so that \eq{sd} is written for the
hermitean matrices. This difference should disappear, however, as
$N_c\ra\infty$ which can be easily proven for the even potential
when $\langle \fr 1N \tr{V^\p(\Phi_x)} \rangle $ vanishes.

The path $C_{xy}$ on the r.h.s.\ of \eq{sd} is always
closed due to the presence of the delta-function. The explicit equation for
the vanishing  contour $C_{xx}=0$ at large $N_c$, when
the factorization holds, reads
\be
\int_{C_1}\frac {d\omega}{2\pi i} \,
\frac{V^\p(\omega)}{\l-\om} E_{\omega}
-2D G_{\l}(1) =E_\l^2
\label{sd0}
\ee
where
\be
E_\l \equiv \left\langle
\frac{1}{N_c}\tr{}\Big( \frac{1}{\l-\Phi_x} \Big) \right\rangle
=\frac{1}{\l}( G_\l(0)+1)
\label{defE}
\ee
with $G_\l$ defined by \eq{sG}.
I have denoted the one-link average by
\be
G_{\l}(1)=G_{\l}(C_{(x\pm\mu)x})
\label{G(1)}
\ee
since the r.h.s.\ does not depend on $x$ and $\mu$ due to the
invariance under translations by a multiple of the lattice spacing and/or
rotations by a multiple of $\pi/2$ on the lattice.
The contour $C_1$ encircles singularities of ${E}_{\omega}$ so that the
integration over $\omega$ on the l.h.s.\ of \eq{sd0} plays the role of a
projector picking up negative powers of $\l$.

\newsection{Loop equations at large $N_c$}

The loop equations for non-vanishing contours $C_{xy}\neq0$
are drastically simplified at $N_c=\infty$ in the strong
coupling region where the closed adjoint Wilson loops~\rf{adjloop} vanish
except the {\it contractable\/} loops (\ie those with vanishing minimal area
$A_{min}(C)$ which are equivalent to $C_{xx}=0$ due to the unitarity of
$U$'s):
\be
W_A(C)=\delta_{0A_{min}(C)}+{\cal O}\left({1\over N_c^2} \right)\,.
\label{1overN}
\ee
While the averages of a new kind
arise on the r.h.s.\ of \eq{sd} for $C_{xx}\neq0$, they obey at
$N_c=\infty$ the following analogue of \eq{1overN}
\be
\left\langle
\frac{1}{N_c}\tr{}{\Big(U(C_{xx})\frac{1}{\l-\Phi_x}\Big)}
\frac{1}{N_c}\tr{}{\Big(U^\dagger(C_{xx})
\frac{1}{\l-\Phi_x}\Big)}\right\rangle  =
\delta_{0,A_{min}(C)} E_\l^2
+{\cal O}\left({1\over N_c^2} \right)
\label{1overNp}
\ee
\ie vanish for $C_{xx}\neq0$.

Hence, the strong coupling loop equation for $C_{xy}\neq0$ at $N_c=\infty$
reads
\bea
 \left\langle \frac{1}{N_c} \tr{}\Big(V^\p(\Phi_x)
U(C_{xy}) \frac{1}{\l-\Phi_y}
U^\dagger(C_{xy}) \Big) \right\rangle
-\sum_{\mu=-D\atop \mu\neq0}^D G_\l(C_{(x+\mu)x}C_{xy})=0
\label{sd1}
\eea
independently of whether $C_{xy}$ is closed or open.

Therefore,
the r.h.s.\ of the loop equation in nonvanishing at $N_c=\infty$ only for
$C_{xy}=0$ (modulo backtrackings) when the proper equation is given
by \eq{sd0}. This property of the strong coupling loop equations at
$N_c=\infty$ allows to find a simple solution.

\newsection{The strong coupling solution}

Let us solve the set~\rf{sd0}, \rf{sd1} of the $N_c=\infty$ loop equations
at strong coupling by the following ansatz in the case of the even potential
\bea
\left\langle \frac{1}{N_c}\tr{}\Big( F(\Phi_x) U_\mu(x)
\Phi_{x+\mu} U^\dagger_\mu(x) \Big) \right\rangle  =
\left\langle \frac{1}{N_c}\tr{}\Big( F(\Phi_x)
\Phi_x \L(\Phi_x) \Big) \right\rangle
\label{Lansatz}
\eea
where  $F(\Phi)$ is arbitrary.
The function $\L(\omega)$ is analytic at $\om=0$:
\be
\L(\om)=\sum_{k=0}^\infty \L_k \om^k\,.
\label{L_k}
\ee
For $G_\l(1)$ which is defined by \eq{G(1)}, \eq{Lansatz} can be written as
\be
G_\l(1)=
\int_{C_1}\frac {d\omega}{2\pi i}\,
\frac{ \om \L(\omega)}{\l-\om}\, E_{\omega}
\label{Fansatz}
\ee
where the contour $C_1$ encircles singularities of $E_\om$, \ie the same
as in \eq{sd0}.

The formula~\rf{Fansatz} extends
to the general potential the one
\be
G_\l(1) = \Lambda_0 \l E_\l \hbox{ \ \ \ \ \ \ \ (quadratic potential)}
\label{qansatz}
\ee
for the quadratic potential
which is associated with~\cite{Mak92}
\be
\L(\om)=\L_0  \hbox{ \ \ \ \ \ \ \ (quadratic potential)}\,,
\label{L0}
\ee
\ie $\L_k=0$ for $k\geq1$.

The substitution of the ansatz~\rf{Fansatz} into \eq{sd0} yields
\be
\int_{C_1}\frac {d\omega}{2\pi i}\,
\frac{\tilde{V}^\p(\omega)}{\l-\om} \,E_{\omega} =E_\l^2
\label{onematrix}
\ee
where
\be
\tilde{V}^\p(\om)=V^\p(\om)-2D\om\L(\om)
\label{tildeV}
\ee
which {\it coincides with the loop equation for the hermitean one-matrix model}
with the potential $\tilde{V}$.
\footnote{For a review, see Ref.~\cite{Mak91}. \label{f2}}

The dependence of $\L(\om)$ on the potential $V$ can be determined from
\eq{sd1}. The simplest way to do this is to consider \eq{sd1} in the case when
$C_{xy}$ is just one link $(x,\mu_0)$ and to take the $1/\l^2$ term of the
$1/\l$ expansion. The resulting equation reads explicitly
\bea
& &\left\langle \frac{1}{N_c} \tr{}\Big(V^\p(\Phi_{x})
U_{\mu_0}(x)\Phi_{x+\mu_0}U^\dagger_{\mu_0}(x) \Big) \right\rangle
- \left\langle \frac{1}{N_c} \tr{}\Phi^2_{x+\mu_0} \right\rangle
\nonumber \\* & &
-\sum_{\mu=-D\atop {\mu\neq0 \atop \mu\neq\mu_0}}^D
 \left\langle \frac{1}{N_c} \tr{}\Big(\Phi_{x-\mu}
U_\mu(x-\mu) U_{\mu_0}(x)\Phi_{x+\mu_0}U^\dagger_{\mu_0}(x)
U^\dagger_\mu(x-\mu) \Big) \right\rangle = 0
\label{sd1(1)}
\eea
which reduces after the substitution of the ansatz~\rf{Fansatz} to
\be
\int_{C_1}\frac {d\omega}{2\pi i}\,\left(\frac
{V^\p(\om)}{\om}\L(\om)- 1-(2D-1)\L^{2}(\om)\right)
 \, \om^2  E_{\omega}=0\,.
\label{nul}
\ee
This equation is satisfied provided that $\L(\om)$ obeys the
quadratic equation
\be
\frac{V^\p(\om)}{\om}-\frac{1}{\L(\om)}-(2D-1){\L(\om)}=0
\label{quadrequation}
\ee
for the generic potential
\be
V(\om)=\sum_{n=1}^\infty t_{2n} \om^{2n}
\label{pott}
\ee
and $t_2\equiv m_0^2$.

The solution to \eq{quadrequation} reads
\be
\L(\om)=\frac{2}{\frac{V^\p(\om)}{\om}+
\sqrt{\left(\frac{V^\p(\om)}{\om}\right)^2+4(1-2D)}}
\label{quadrsolution}
\ee
which recovers the one~\cite{Mak92} for the quadratic potential
when
\be
V^\p(\om)=2 m_0^2 \om\hbox{ \ \ \ \ \ \ \  (quadratic potential)}
\ee
and \eq{L0} holds.
The fact that the ansatz~\rf{Fansatz} with $\L(\om)$ given by
\eq{quadrsolution} satisfies \eq{sd1} means that it is
indeed a solution providing \eq{onematrix} with
\be
 \frac{\tilde{V}^\p(\om)}{\om}=
\frac{D-1}{2D-1}\,\frac{V^\p(\om)}{\om}+\frac{D}{2D-1}
\sqrt{\left(\frac{V^\p(\om)}{\om}\right)^2+4(1-2D)}
\label{tildeVp}
\ee
is satisfied.

The one-cut solution to this equation for an arbitrary potential
is well-known~\cite{Mig83}
\be
E_\l=\int_{C_1}\frac {d\omega}{4\pi i}\,
\frac{\tilde{V}^\p(\omega)}{\l-\om}
\sqrt{\frac{(\l-x)(\l-y)}{(\om-x)(\om-y)}}
\label{zero}
\ee
where $x$ and $y$ are expressed via $\tilde{V}$ by
\be
\int_{C_1}\frac {d\omega}{2\pi i}\,
\frac{\tilde{V}^\p(\om)}{\sqrt{(\om-x)(\om-y)}}=0\,,
\hspace{1.0cm} \int_{C_1}\frac {d\omega}{2\pi i}\,
 \frac{\om \tilde{V}^\p(\om)}{\sqrt{(\om-x)(\om-y)}}=2\,.
 \label{xandy}
\ee
One gets $x=-y$ for the even potential.

The formulas~\rf{Fansatz}, \rf{tildeVp}, \rf{zero} and \rf{xandy}  completes
the solution of the strong coupling loop equations of KMM at $N_c=\infty$.
The saddle point value of $\Phi_x$ is totally determined to be $x$-independent
and  is described, modulo a gauge transformation, by the spectral density
\be
\rho(\l)=\frac {1}{2\pi^2}\int_y^x dt\,
\frac{\tilde{V}^\p(t)-\tilde{V}^\p(\l)}{t-\l}
\sqrt{\frac{(x-\l)(\l-y)}{(x-t)(t-y)}} \hbox{ \ \ for \ \ } y<\l<x
\label{rho}
\ee
with support $y<\l<x$.
Properties of this solution are discussed in the next section.

\newsection{Relation to the large mass expansion}

The peculiar form~\rf{Fansatz} of
the strong coupling solution at $N_c=\infty$ can
be understood in the framework of the large mass expansion.
To this aim let us consider the one-link correlator
of the gauge fields
\be
\left\langle \frac {1}{N_c} \tr{} \Big(t^a U \Phi_{x+\mu}
U^\dagger\Big)\right\rangle_{U} \equiv
\frac{\int dU\,\e^{N_c\tr{}\Big( \Phi_x U
\Phi_{x+\mu} U^\dagger\Big)} \frac {1}{N_c} \tr{} \Big(t^aU
\Phi_{x+\mu} U^\dagger\Big)} {\int dU\,\e^{N_c
\tr{}\Big(\Phi_x U \Phi_{x+\mu} U^\dagger\Big)}}
\label{onelink}
\ee
where the averaging is only w.r.t.\ $U$ while $\Phi_x$ and
$\Phi_{x+\mu}$ play the role of external fields.
$t^a$ ($a=1,\ldots,N_c^2-1$) stand for generators of $SU(N_c)$ which
are normalized by
\be
\frac{1}{N_c}\tr{}t^at^b = \delta^{ab} \,.
\ee
As was proposed in Refs.~\cite{Mig92a,Mig92d},
the following formula holds at $N_c=\infty$:
\be
\left\langle \frac {1}{N_c} \tr{}\Big( t^aU \Phi_{x+\mu}
U^\dagger
\Big)\right\rangle_{U} =\sum_{m=0}^\infty  \Lambda_{m}
\frac{1}{N_c}\tr{}\left(t^a\Phi^{m+1}_x\right)\,.
\label{Lambda}
\ee
It is shown in Appendix~B how this formula can be obtained
in the large mass expansion.

\eq{Lambda} allows to explain the solution of the previous section
as follows. Let us multiply both sides of~\eq{Lambda} by
\/ tr$\left(t^a F(\Phi_x)\right)$ which gives, using the
completeness condition~\rf{completeness},
\bea
 & & \frac{1}{N_c}\tr{}\Big( F(\Phi_x) U
\Phi_{x+\mu} U^\dagger \Big)  = \nonumber \\*
& & \frac{1}{N_c}\tr{}\Big( F(\Phi_x)
\Phi_x \L(\Phi_x) \Big) +
 \frac{1}{N_c}\tr{} \Big( F(\Phi_x) \Big)
 \frac{1}{N_c}\tr{} \Big(\Phi_{x+\mu}-\Phi_x \L(\Phi_x)\Big)
\label{ansatzp}
\eea
where ${\Lambda}(\Phi)$ is defined by \eq{L_k}.
The second term on the r.h.s.\ which is due to the difference
between the adjoint representation and the hermitean matrices
vanishes for the even potential at $N_c=\infty$. Hence,
\eq{ansatzp} for $\Phi_{x}$ and $\Phi_{x+\mu}$ given by the saddle
point matrix $\Phi_S$ recovers \eq{Lansatz}. On the other hand,
the solution of the previous section allows to calculate $\L_m$ in
\eq{Lambda} as the coefficients on the expansion
of~\rf{quadrsolution} in $\om$.

\newsection{Discussion}

The above strong coupling solution is realized at given $D$ only in some region
of the couplings $t_{2k}$'s
entering the potential~\rf{pott}. At $D=0$ it coincides
with the well-known solution of the hermitean one-matrix model. At any $D$ but
$t_2\neq0$, $t_{2k}=0$ for $k\geq1$, the solution coincides with the one for
the
quadratic potential~\cite{Gro92}. For this reason I expect that it is realized
in some region around this point.  The condition is that \eq{xandy}
should yield real
$x$ and $y$ and the spectral density~\rf{rho}, which describes the distribution
of eigenvalues of the saddle point matrix $\Phi_S$,
should be positive.  This is a
restriction on the one-cut solution which is satisfied in some region of values
of the couplings $t_{2k}$'s. It is well known that for the simple quartic
potential
\be
\tilde{V}(\om)=\tilde{t}_2 \om^2 + \tilde{t}_4 \om^4
\ee
the one-cut spectral density is positive for
$|\tilde{t_4}|\leq \tilde{t}_2^2/12$. This is,
however, a nontrivial restriction on the potential $V$ since
\be
V^\p(\om)=D \sqrt{(\tilde{V}^\p(\om))^2+4\om^2} -(D-1)\tilde{V}^\p(\om)\,.
\ee

One more restriction on the one-cut solution is given by the requirement that
the expession under the square root
in \eq{tildeVp} must be positive for any $\om$ which belongs
to the support of the spectral density. If this expression becomes negative for
some values of $t_{2k}$'s
this simply means that the one-cut solution is not realized
and one should look for a more sofisticated support (multi-cut solutions).
This is quite standard for the large-$N$ phase transitions which occur
at the values of couplings where the behavior of the spectral density
changes.

It is interesting to compare our solution with that obtained
by the Riemann--Hilbert method.  It
is easy to identify  the function ${\cal T}_\l(z)$ which
determines the solution of Ref.~\cite{Mig92a} with the one-link correlator
\be
{\cal T}_\l(z)-1= \left\langle
\frac {1}{N_c} \tr{}{\Big( \frac{1}{z-\Phi_S} U
\frac{1}{\l- \Phi_S}
U^\dagger \Big)} \right\rangle_{U}
\label{Tlz}
\ee
which is defined by the same average as in \eq{onelink}.
The proof is based on the following extension of the Migdal procedure.
Let us define~\cite{Mig92a}
\be
G_\l(\Phi_x) \equiv \frac {1}{I(\Phi_x, \Phi_{x+\mu})}\,
\frac{1}{\l-\frac{1}{N_c}\frac{\d}{\d\Phi_x}}\, I(\Phi_x, \Phi_{x+\mu})
\ee
where $I$ stands for the Itzykson--Zuber integral
\be
I(\Phi_x, \Phi_{x+\mu})=\int dU\,\e^{N_c
\tr{}\left(\Phi_x U \Phi_{x+\mu} U^\dagger \right)} \,.
\label{IZ}
\ee
By a direct differentiation of \eq{IZ} one gets
\be
\frac {1}{N_c} \tr{}\Big( \frac{1}{z-\Phi_x}G_\l(\Phi_x)\Big)= \left\langle
\frac {1}{N_c} \tr{}{\Big( \frac{1}{z-\Phi_x} U
\frac{1}{\l- \Phi_{x+\mu}}
U^\dagger \Big)} \right\rangle_{U}\,.
\label{Glz}
\ee
Rewriting the l.h.s.\ via the spectral density,
substituting for $\Phi_x$ and $\Phi_{x+\mu}$ the saddle point value $\Phi_S$
and remembering the definition
of ${\cal T}_\l(z)$~\cite{Mig92a}
\be
{\cal T}_\l(z) = 1+ \int d\mu \frac{\rho(\mu)G_\l(\mu)}{\mu-\l}\,,
\ee
one proves \eq{Tlz}.

{}From \eq{Glz} it is easy to see alternatively that $N_c=\infty$
\be
{\cal T}_\l(z)-1= \left\langle
\frac {1}{N_c} \tr{}{\Big( \frac{1}{z-\Phi_x} U_\mu(x)
\frac{1}{\l- \Phi_{x+\mu}}
U^\dagger_\mu(x) \Big)} \right\rangle
\label{Tlzp}
\ee
where the average is w.r.t.\ the same measure as in \eq{spartition}.
Therefore, we get asymptotically
\be
{\cal T}_\l(z) =1 + \frac{E_\l}{z} + \frac{G_\l(1)}{z^2} + \ldots
\hbox{ \ \ \ \ \ \ as \ } z\ra\infty
\label{ourT}
\ee
so that our $G_\l(1)$ is to be compared with the ${\cal O}(z^{-2})$ term in
the expansion of ${\cal T}_\l(z)$. It would be very interesting to
calculate exactly the correlator on the r.h.s.\ of \eq{Tlzp},
which should be manifestly symmetric w.r.t.\ $\l$ and $z$, by solving the loop
equations in order to compare
with ${\cal T}_\l(z)$ of Ref.~\cite{Mig92a}.

It is worth mentioning that in finding the solution I did not use the fact
that $\Phi_x$ can be diagonalized. For this reason
a solution analogous to that of this paper exists
for the adjoint fermion model~\cite{KhM92b} which reduces to the {\it
complex} one-matrix model~\cite{Mak91}. This result, as well as an
analysis of the $D\leq1$ case, will be published elsewhere.
The most interesting
question is whether the solution of this paper admits a continuum limit for
$D>1$.

\subsection*{Acknowledgements}

I thank the theoretical physics department of UAM for the hospitality in Madrid
where a part of the work was done.
This paper was not supported by grands of APS or IFS.

\eop

\setcounter{section}{0}
\appendix{Derivation of the loop equations}

Let us consider an equation which results from the invariance of the measure
over $\Phi$ in the open-loop average~\rf{sG} under an infinitesimal shift
\be
\Phi_x\ra\Phi_x+\xi_x
\label{xi}
\ee
of $\Phi_x$ at the given site $x$ with $\xi_x$ being an infinitesimal
hermitean matrix. For KMM one should impose
$\tr{\xi_x}=0$ in order for the shifted matrix to belong to the adjoint
representation of $SU(N_c)$. Since this condition should be inessential
as $N_c\ra\infty$, I derive loop equations for the hermitean model,
defined by the partition function~\rf{spartition} with the integration
going over arbitrary hermitean matrices $\Phi_x$. $\xi_x$ in \eq{xi} is then
arbitrary hermitean.

It is convenient to introduce
$N_c^2$ generators
\be
[t^A]_{ij}=\Big( \delta_{ij},\, [t^a]_{ij} \Big)
\label{t^A}
\ee
with $t^a$ ($a=1,\ldots,N_c^2-1$) being the standard generators of
$SU(N_c)$.  The generators~\rf{t^A} obey the following normalization
\be
\frac{1}{N_c} \tr{t^A t^B}=\delta^{AB}
\label{normalization}
\ee
and completeness condition
\be
[t^A]_{ij}[t^A]_{kl}=N_c \delta_{il} \delta_{kj} \,.
\label{completeness}
\ee
An arbitrary $N_c\times N_c$ hermitean matrix $\Phi$ can be represented as
\be
\Phi=t^A \Phi^A \hbox{ \ \ \ \ where \ \ \ }
\Phi^A=\Big( \frac{1}{N_c} \tr{\Phi},\, \frac{1}{N_c} \tr{t^a\Phi} \Big)
\ee
with $\Phi^0=\fr{1}{N_c}\tr{\Phi}$ vanishing if $\Phi$ is taken in the
adjoint representation of $SU(N_c)$.

To derive the loop equation I apply a trick similar to that used in deriving
loop equations of QCD~\cite{Mig83}. Let us consider the loop average
\be
\left\langle \frac{1}{N_c} \tr{\Big(t^A U(C_{xy}) \frac{1}{\l-\Phi_y}
U^\dagger(C_{xy})\Big)}  \right\rangle =0\,,
\label{vanishing} \ee
where the averaging is taken with the same measure as in \eq{spartition},
which vanishes due to the gauge invariance. Performing the
shift~\rf{xi} of $\Phi_x$, using the invariance of the measure and
calculating $\d/\d \Phi^B(x)$, one gets
\bea
\left\langle \frac{1}{N_c} \tr{\Big(t^B V^\p(\Phi_x)\Big)}
\frac{1}{N_c} \tr{\Big(t^A U(C_{xy}) \frac{1}{\l-\Phi_y}
U^\dagger(C_{xy})\Big)} \right\rangle  \nonumber \\*
-\sum_{\mu=-D\atop\mu\neq0}^D
\left\langle \frac{1}{N_c} \tr{}\Big(
t^B U_\mu(x)\Phi_{x+\mu}U_\mu^\dagger(x)
\Big) \frac{1}{N_c} \tr{t^A U(C_{xy})
\frac{1}{\l-\Phi_y} U^\dagger(C_{xy})} \right\rangle \nonumber \\*
= \delta_{xy}
\left\langle \frac{1}{N_c^3} \tr{\Big(t^A U(C_{xy})\frac{1}{\l-\Phi_y}
t^B \frac{1}{\l-\Phi_y} U^\dagger(C_{xy})\Big)}
\right\rangle \;.
\label{AB}
\eea
The l.h.s.\ of this equation results from the variation of the action while
the r.h.s.\ represents the commutator term resulting from the variation of
the integrand.

The averaging over the gauge group picks up two nonvanising invariant
equations for the hermitean matrices. The first one can be obtained
contracting \eq{AB} by $\delta^{AB}$ ($A,B=0,\ldots,N_c^2-1$)
while the second one is given by the $A,B=0$ component.

The first equation reads
\bea
 \left\langle \frac{1}{N_c} \tr{\Big(V^\p(\Phi_x)
U(C_{xy}) \frac{1}{\l-\Phi_y}
U^\dagger(C_{xy})\Big)} \right\rangle  \nonumber \\*
-\sum_{\mu=-D\atop\mu\neq0}^D
\left\langle \frac{1}{N_c} \tr{\Big( \Phi_{x+\mu}
U(C_{(x+\mu)x}C_{xy})
 \frac{1}{\l-\Phi_y}U^\dagger(C_{(x+\mu)x}C_{xy})}  \Big)
\right\rangle \nonumber \\* = \delta_{xy}
\left\langle \frac{1}{N_c} \tr{\Big( U(C_{xy})\frac{1}{\l-\Phi_y}\Big)}
\frac{1}{N_c} \tr{\Big( \frac{1}{\l-\Phi_y}
U^\dagger(C_{xy})\Big)}
\right\rangle
\label{AA}
\eea
where the contour $C_{(x+\mu)x}C_{xy}$ is obtained by attaching the link
$(x,\mu)$ to the path $C_{xy}$ at the end point $x$ as is depicted
in Fig.~\ref{fig.4}.
Using the definition~\rf{sG}, this equation can be
written finally in the form~\rf{sd}.

The second equation which is given by the $A,B=0$ component of \eq{AB} reads
\bea
 \left\langle \frac{1}{N_c} \tr{V^\p(\Phi_x )}
\frac{1}{N_c}\tr{\frac{1}{\l-\Phi_y}} \right\rangle
-\sum_{\mu=-D\atop\mu\neq0}^D
\left\langle \frac{1}{N_c} \tr{ \Phi_{x+\mu}}
 \frac{1}{N_c} \tr{ \frac{1}{\l-\Phi_y}}
\right\rangle \nonumber \\* = \left\langle
\frac{1}{N_c^3}\tr{\frac{1}{\Big( \l-\Phi_y\Big)^2}}\right\rangle\,.
\label{00}
\eea
In the large-$N_c$ limit when the factorization holds, \eq{00} is
automatically satisfied as a consequence of the ${\cal O}(\l^{-1})$ term in
\eq{AA}.

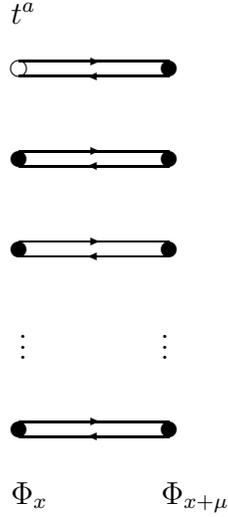
\begin{figure}[tbp]
\unitlength=1.00mm
\linethickness{0.5pt}
\begin{picture}(33.00,77.00)(-70,40)
\put(20.00,105.00){\vector(-1,0){11.00}}
\put(9.00,105.00){\line(-1,0){9.00}}
\put(0.00,106.00){\circle{2.00}}
\put(0.00,107.00){\vector(1,0){11.00}}
\put(11.00,107.00){\line(1,0){9.00}}
\put(20.00,106.00){\circle*{2.00}}
\put(0.00,95.00){\vector(1,0){11.00}}
\put(11.00,95.00){\line(1,0){9.00}}
\put(0.00,94.00){\circle*{2.00}}
\put(20.00,94.00){\circle*{2.00}}
\put(20.00,93.00){\vector(-1,0){11.00}}
\put(9.00,93.00){\line(-1,0){9.00}}
\put(0.00,83.00){\vector(1,0){11.00}}
\put(11.00,83.00){\line(1,0){9.00}}
\put(0.00,82.00){\circle*{2.00}}
\put(20.00,82.00){\circle*{2.00}}
\put(20.00,81.00){\vector(-1,0){11.00}}
\put(9.00,81.00){\line(-1,0){9.00}}
\put(0.00,59.00){\vector(1,0){11.00}}
\put(11.00,59.00){\line(1,0){9.00}}
\put(0.00,58.00){\circle*{2.00}}
\put(20.00,58.00){\circle*{2.00}}
\put(20.00,57.00){\vector(-1,0){11.00}}
\put(9.00,57.00){\line(-1,0){9.00}}
\put(-1,112){\makebox(0,0)[lb]{$t^a$}}
\put(-1,51){\makebox(0,0)[lt]{$\Phi_x$}}
\put(19,51){\makebox(0,0)[lt]{$\Phi_{x+\mu}$}}
\put(0,70){\makebox(0,0)[lc]{$\vdots$}}
\put(20,70){\makebox(0,0)[rc]{$\vdots$}}
\end{picture}
\caption[x]   {\hspace{0.2cm}\parbox[t]{13cm}
{\small
   The graphic representation of the large mass expansion of the one-link
   correlator~\rf{onelink}. The right filled circles
   represent $\Phi_{x+\mu}$
   while the left ones represent $\Phi_x$. The empty circle represents $t^a$.}}
   \label{fig.6}
\end{figure}

\appendix{The one-link correlator at large $N_c$}

\eq{Lambda} can be derived
for $\Phi$ given by the saddle point (master field) configuration
analyzing the large mass expansion which allows to calculate
the one-link correlator~\rf{onelink}  in the
strong coupling phase (\ie before an expected large-$N_c$ phase transition).
To calculate it, let us expand the numerator in powers of $\Phi$ as is
depicted in Fig.~\ref{fig.6}.

The idea is now {\it not\/} to calculate the complicated integral over
$U_\mu(x)$ in \eq{onelink} but rather substitute for $\Phi_x$ and
$\Phi_{x+\mu}$ the saddle point value $\Phi_S$ which is determined by the
future integration over $\Phi$ according to \eq{spartition}. Notice, that
the integrals over $\Phi_x$ and $\Phi_{x+\mu}$ are independent to each order of
the large mass expansion. Therefore, one substitutes
\be
(\Phi_x^a)_S (\Phi_x^b)_S = K \frac{1}{N_c^2}\delta^{ab}
\label{K}
\ee
and
\be
(\Phi_x^a)_S (\Phi_{x+\mu}^b)_S =0
\label{KK}
\ee
where
\be
K=\frac{1}{N_c} \tr{}\Phi_S^2
\ee
and~\rf{KK} vanishes due to the gauge invariance.

After the use of Eqs.~\rf{K}, \rf{KK} and the completeness
condition~\rf{completeness} the color indices are contracted in such
a way that all the matrices $U_\mu(x)$ disappear due to the unitarity.  The
simplest contractions are depicted in Fig.~\ref{fig.7}.
\begin{figure}[htb]
\unitlength=1.00mm
\linethickness{0.4pt}
\begin{picture}(33.00,58.00)(-5,64)
\put(20.00,100.00){\oval(10.00,10.00)[r]}
\put(20.00,100.00){\oval(14.00,14.00)[r]}
\put(20.00,105.00){\vector(-1,0){11.00}}
\put(20.00,93.00){\vector(-1,0){11.00}}
\put(9.00,93.00){\line(-1,0){9.00}}
\put(9.00,105.00){\line(-1,0){9.00}}
\put(0.00,107.00){\vector(1,0){11.00}}
\put(11.00,107.00){\line(1,0){9.00}}
\put(0.00,95.00){\vector(1,0){11.00}}
\put(11.00,95.00){\line(1,0){9.00}}
\put(0.00,106.00){\circle{2.00}}
\put(0.00,94.00){\circle*{2.00}}
\put(33.00,100.00){\makebox(0,0)[lc]{$=\;\;K\;\frac{1}{N_c}\;
				\hbox{tr}\,t^a\,\Phi_x$}}
\put(13,71){\makebox(0,0)[cc]{a)}}
\put(-1,112){\makebox(0,0)[lb]{$t^a$}}
\put(-1,87){\makebox(0,0)[lt]{$\Phi_x$}}
\end{picture}
\unitlength=1.00mm
\linethickness{0.5pt}
\begin{picture}(33.00,58.00)(-50,58)
\put(20.00,100.00){\oval(10.00,10.00)[r]}
\put(22.00,100.50){\oval(9.00,13.00)[r]}
\put(20.00,88.00){\oval(10.00,10.00)[r]}
\put(22.00,87.50){\oval(9.00,13.00)[r]}
\put(20.00,105.00){\vector(-1,0){11.00}}
\put(20.00,93.00){\vector(-1,0){11.00}}
\put(9.00,93.00){\line(-1,0){9.00}}
\put(9.00,105.00){\line(-1,0){9.00}}
\put(0.00,107.00){\vector(1,0){11.00}}
\put(11.00,107.00){\line(1,0){11.00}}
\put(0.00,95.00){\vector(1,0){11.00}}
\put(11.00,95.00){\line(1,0){9.00}}
\put(0.00,83.00){\vector(1,0){11.00}}
\put(11.00,83.00){\line(1,0){9.00}}
\put(22.00,81.00){\vector(-1,0){13.00}}
\put(9.00,81.00){\line(-1,0){9.00}}
\put(0.00,82.00){\circle*{2.00}}
\put(0.00,106.00){\circle{2.00}}
\put(0.00,94.00){\circle*{2.00}}
\put(33.00,94.00){\makebox(0,0)[lc]{$=\;\;
    t_3\, K^3 \;\frac{1}{N_c}\; \hbox{tr}\,t^a\,\Phi_x^2$}}
\put(13,65){\makebox(0,0)[cc]{b)}}
\put(-1,112){\makebox(0,0)[lb]{$t^a$}}
\put(-1,74){\makebox(0,0)[lt]{$\Phi_x$}}
\end{picture}
\caption[x]   {\hspace{0.2cm}\parbox[t]{13cm}
{\small
   The contraction of color indices for diagrams of Fig.~\ref{fig.6}. \\
   a) The diagram results in the contribution to ${\Lambda}_0$. \\
   b) The diagram results in the contribution to ${\Lambda}_{1}$.
   }}
\label{fig.7}
\end{figure}
Only the connected diagrams should be taken into account due the
presence of the denominator.
The diagrams of the type of Fig.~\ref{fig.7}a are the only ones which
emerge for the quadratic potential. They always result in
${\Lambda}_0$. The diagrams of the type Fig.~\ref{fig.7}b appear when the
interaction is present. They result in
${\Lambda}_m$ with $m\geq 1$.

Finally, let us notice that this structure of ${\Lambda}_m$ is not spoiled by
the fact that $\Phi_x$ enters $2D$ links emanating from the point $x$. This
affect only combinatorics making $K$ to be $D$-dependent.

\eop

\end{document}